\documentclass[10pt,twocolumn,letterpaper]{article}

\usepackage{cvpr}              

\usepackage{multirow}
\usepackage{amssymb}
\usepackage{bbding}

\definecolor{cvprblue}{rgb}{0.21,0.49,0.74}
\usepackage[pagebackref,breaklinks,colorlinks,allcolors=cvprblue]{hyperref}

\def\paperID{14960} 
\def\confName{CVPR}
\def\confYear{2026}

\title{OmniSonic: Towards Universal and Holistic Audio Generation \\ from Video and Text}

\author{Weiguo Pian\textsuperscript{\rm 1} \quad Saksham Singh Kushwaha\textsuperscript{\rm 1} \quad Zhimin Chen\textsuperscript{\rm 2} \quad Shijian Deng\textsuperscript{\rm 1} \\
\quad Kai Wang\textsuperscript{\rm 3} \quad Yunhui Guo\textsuperscript{\rm 1} \quad Yapeng Tian\textsuperscript{\rm 1} \\
\textsuperscript{\rm 1} The University of Texas at Dallas \quad \textsuperscript{\rm 2} Clemson University \quad \textsuperscript{\rm 3} University of Toronto \\
}

\begin{document}
\maketitle

\begin{abstract}

In this paper, we propose Universal Holistic Audio Generation (UniHAGen), a task for synthesizing comprehensive auditory scenes that include both on-screen and off-screen sounds across diverse domains (e.g., ambient events, musical instruments, and human speech). Prior video-conditioned audio generation models typically focus on producing on-screen environmental sounds that correspond to visible sounding events, neglecting off-screen auditory events. While recent holistic joint text-video-to-audio generation models aim to produce auditory scenes with both on- and off-screen sound but they are limited to non-speech sounds, lacking the ability to generate or integrate human speech. To overcome these limitations, we introduce OmniSonic, a flow-matching–based diffusion framework jointly conditioned on video and text. It features a TriAttn-DiT architecture that performs three cross-attention operations to process on-screen environmental sound, off-screen environmental sound, and speech conditions simultaneously, with a Mixture-of-Experts (MoE) gating mechanism that adaptively balances their contributions during generation. Furthermore, we construct UniHAGen-Bench, a new benchmark with over one thousand samples covering three representative on/off-screen speech–environment scenarios. Extensive experiments show that OmniSonic consistently outperforms state-of-the-art approaches on both objective metrics and human evaluations, establishing a strong baseline for universal and holistic audio generation. Project page: \url{https://weiguopian.github.io/OmniSonic_webpage/}

\end{abstract}    
\section{Introduction}
\label{sec:intro}

\begin{figure}
    \centering
    \includegraphics[width=0.48\textwidth]{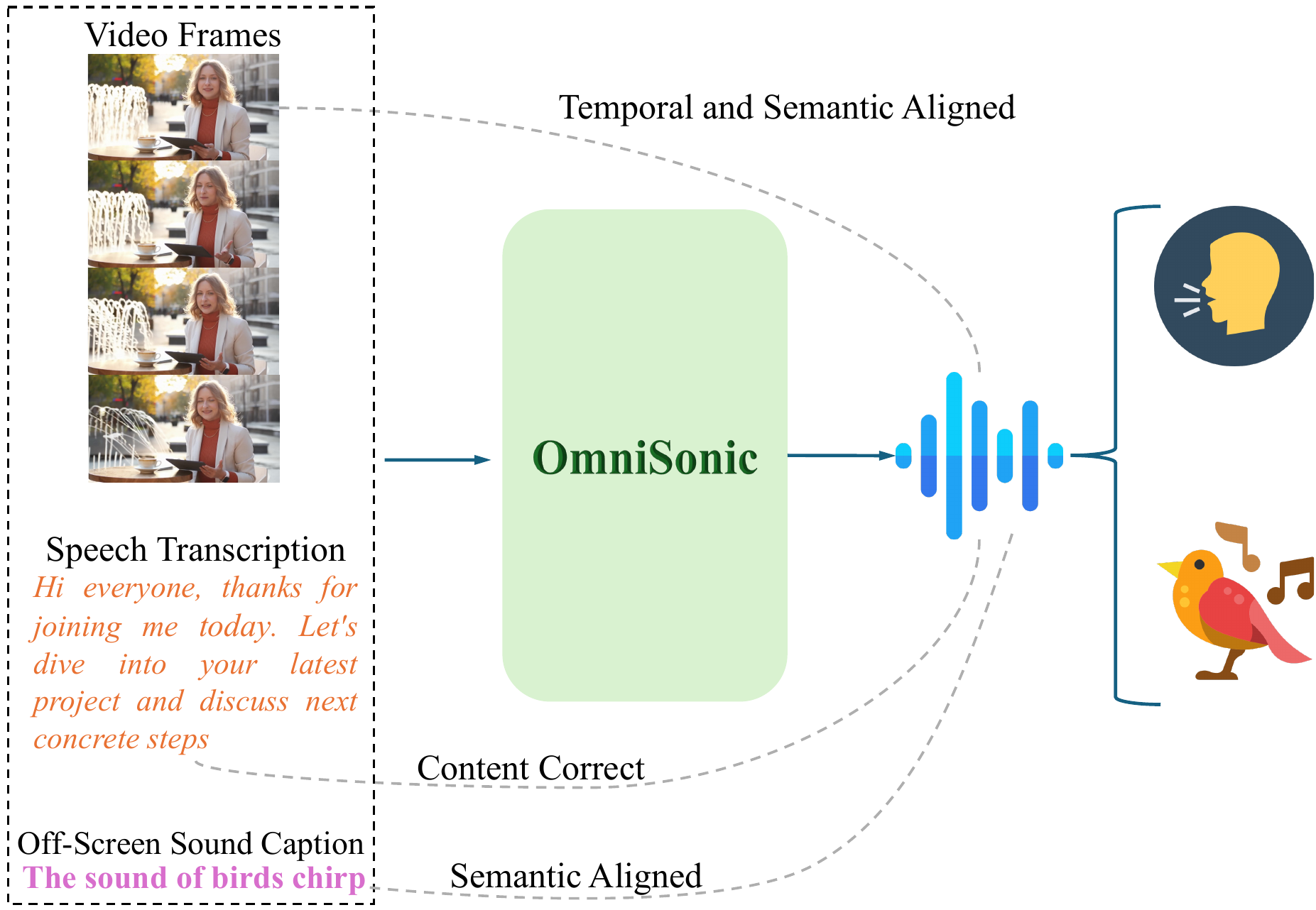}
    \vspace{-15pt}
    \caption{Illustration of the proposed Universal Holistic Audio Generation (UniHAGen) task. The example depicts the scenario of on-screen speech with off-screen environmental sound. Our model is able to generate audio that is temporally and semantically aligned with the video, consistent with the environmental captions, and faithful to the given speech transcription.}
    \label{fig:teaser}
    \vspace{-15pt}
\end{figure}

Imagine a silent video of a person speaking while birds chirp off screen. To render this scene realistically, a video-to-audio generation system must synthesize the speaker’s voice synchronized with lip movements and generate off-screen environmental sounds that provide context and atmosphere. Achieving such auditory scene synthesis, which covers visible and invisible sources across diverse sound domains (\eg, speech, ambient sound, music), remains a significant challenge for current generative models.


With recent advances in diffusion models~\cite{ho2020denoising,rombach2022high,lipman2023flow,liu2023flow}, audio generation has made remarkable progress in producing high-fidelity and diverse sounds~\cite{luo2023diff,liu2023audioldm,liu2024audioldm,ghosal2023text,majumder2024tango,zhang2024foleycrafter,shan2025hunyuanvideo,cheng2025mmaudio,wang2024frieren,chen2025video,li2025tri}. These models iteratively refine noise into realistic audio waveforms through denoising processes, enabling precise control over acoustic details and semantic alignment with conditioning inputs such as video, text, or both. Despite these advances, video-conditioned audio generation models still face two major shortcomings in achieving universal and holistic sound generation.
First, conventional video-conditioned audio generation models are typically limited to producing on-screen sounds~\cite{luo2023diff, cheng2025mmaudio, zhang2024foleycrafter,shan2025hunyuanvideo} that correspond to visible actions or objects, overlooking off-screen auditory events that are crucial for creating a complete auditory scene.
Second, most existing models focus primarily on environmental sounds (\textit{e.g.}, ambient noise, musical instruments) and struggle to generalize to other sound types such as human speech or complex mixtures of speech and environmental sounds. These limitations constrain current models from perceiving and generating high-quality audio in complex scenes that may involve diverse sound domains and invisible sound sources.

Recent joint video–text approaches such as VinTAGe~\cite{kushwaha2025vintage} and SonicVisionLM~\cite{xie2024sonicvisionlm} move toward holistic scene synthesis by generating on-screen sounds from video and controlling off-screen sounds with text. However, they remain limited to environmental sound generation and cannot produce high-quality speech. Consequently, they fail to model scenarios where speech and environmental sounds coexist and interact, revealing a core gap in unified cross-domain audio generation. In parallel, environmental speech generation models~\cite{lee2024voiceldm,jung2025voicedit} synthesize speech under predefined acoustic conditions but rely solely on text, lacking the visual grounding needed to distinguish on-screen from off-screen sources. Their simplified and domain-specific assumptions further limit applicability to video-based settings, so these approaches cannot handle diverse, dynamic scenes with both visible and invisible sound sources.


To address these challenges, we introduce Universal Holistic Audio Generation (UniHAGen), a new task aimed at synthesizing \textit{complete auditory scenes} that include both \textit{on-screen} and \textit{off-screen} sounds across \textit{diverse domains} such as ambient noise, musical instruments, and human speech.
Specifically, UniHAGen defines three representative scenarios: 
(1) on-screen environmental sound + off-screen human speech,
(2) on-screen human speech + off-screen environmental sound, and
(3) on-screen environmental sound + off-screen environmental sound + off-screen human speech.
Unlike previous ``holistic" frameworks that focus solely on environmental sounds, UniHAGen unifies speech and environmental sound generation within a single multimodal paradigm, enabling models to reason jointly about heterogeneous sound domains.
Fig.~\ref{fig:teaser} illustrates Scenario 2, where the model must generate on-screen speech consistent with the transcription and visual frames while simultaneously producing off-screen environmental sounds aligned with the accompanying captions.
This example highlights the central challenge of UniHAGen: generating coherent, realistic audio that spans both visible and invisible sound sources across multiple domains.

To tackle the UniHAGen task, we propose OmniSonic, a diffusion-based framework jointly conditioned on text and video inputs. OmniSonic is composed of an environmental text encoder, a speech transcription encoder, a visual encoder, an audio VAE~\cite{kingma2013auto}, and a TriAttn-DiT module. The TriAttn-DiT extends the DiT architecture~\cite{peebles2023scalable} by incorporating triple cross-attention mechanisms, enabling the model to effectively integrate and reason over on-screen environmental captions, off-screen environmental captions, and speech transcriptions for universal holistic audio generation. Moreover, each TriAttn-DiT block incorporates a Mixture-of-Experts (MoE)-based Gating module, which adaptively computes a weighted sum of the three cross-attention outputs, enabling dynamic balancing among on-screen environmental sound caption, off-screen environmental sound caption, and speech transcription information during audio generation. To the best of our knowledge, there is a lack of an appropriate benchmark that aligns with the scenarios defined in our proposed UniHAGen task. To address this gap, we construct a new benchmark, named UniHAGen-Bench, to comprehensively evaluate model performance in the UniHAGen setting. 
Experimental evaluations on UniHAGen-Bench demonstrate that our proposed OmniSonic significantly outperforms state-of-the-art baselines in both objective metrics and subjective human evaluations, highlighting the superiority and effectiveness of our approach. In summary, this paper contributes the following:
\begin{itemize}
    \item We propose UniHAGen, a realistic and comprehensive setting for generating holistic audio that contains both on-screen and off-screen events across diverse domains.
    \item We introduce OmniSonic, a diffusion-based framework jointly conditioned on video and text with a TriAttn-DiT backbone and MoE gating for adaptive fusion.
    \item We construct UniHAGen-Bench, a benchmark covering three representative scenarios for systematic evaluation.
    \item Extensive experiments show that OmniSonic achieves state-of-the-art performance across objective and subjective metrics, validating its effectiveness for the task.
\end{itemize}

\section{Related Works}
\label{sec:related_works}

\textbf{Audio Generation.}
Early audio generation models employed discrete audio representations such as codec tokens for text-to-audio (T2A) synthesis~\cite{kreuk2023audiogen}. With the advent of latent diffusion models, LDM-based T2A approaches~\cite{liu2023audioldm,liu2024audioldm,ghosal2023text,huang2023make} became dominant, substantially improving perceptual quality and diversity through denoising in latent audio spaces. Meanwhile, video-to-audio (V2A) generation has shown strong potential for creating sound for silent videos. Earlier methods based on GANs or autoregressive models~\cite{iashin2021taming,sheffer2023hear,zhou2018visual} were later surpassed by LDM-based frameworks such as Diff-Foley~\cite{luo2023diff} and Frieren~\cite{wang2024frieren}, which improved temporal coherence and acoustic realism. 
Recently, joint video–text-to-audio (VT2A) models such as MMAudio~\cite{cheng2025mmaudio}, MultiFoley~\cite{chen2025video}, and HunyuanVideo-Foley~\cite{shan2025hunyuanvideo} employ multimodal diffusion Transformers~\cite{esser2024scaling} to enhance semantic consistency and synchronization. VinTAGe~\cite{kushwaha2025vintage} introduces video–text conditioning to jointly model on- and off-screen sounds for holistic auditory scene synthesis.
However, these VT2A models remain limited to non-speech domains. In parallel, environmental speech generation systems~\cite{lee2024voiceldm,jung2025voicedit} rely solely on textual input without visual cues, and their predefined acoustic environments are simple and domain-specific. 
This gap motivates us to develop a unified, joint video-and-text conditioned framework for universal holistic audio generation that handles both speech and environmental sounds.

\vspace{2mm}
\noindent
\textbf{Diffusion Models.}
Diffusion models~\cite{ho2020denoising,song2021scorebased} have become a leading paradigm for generative modeling, achieving strong results in vision~\cite{rombach2022high,peebles2023scalable,singer2023makeavideo,ho2022video,ho2022imagen}, audio~\cite{liu2023audioldm,huang2023make}, and beyond.
Early work such as DDPM~\cite{ho2020denoising} formulates data generation as a discrete-time Markov denoising process, gradually transforming Gaussian noise into data samples through learned reverse transitions. Subsequent works generalized this framework to continuous-time formulations using stochastic differential equations or their deterministic ODE counterparts~\cite{song2021scorebased}. Latent diffusion models~\cite{rombach2022high} further enhanced efficiency by performing the diffusion process in the latent space of pretrained autoencoders, enabling high-fidelity and scalable generation. More recent advances such as Flow Matching~\cite{lipman2023flow,liu2023flow} directly learn continuous velocity fields between noise and data distributions, offering more stable training and faster sampling. Building on these developments, OmniSonic adopts a diffusion framework based on flow matching for efficient, temporally coherent, universal holistic audio generation.

\section{Method}
\label{sec:method}

\subsection{Preliminaries}

\textbf{Flow Matching.}
Flow Matching~\cite{lipman2023flow,liu2023flow} is a generative modeling framework that learns a continuous-time vector field to transform samples from a simple prior distribution (\textit{e.g.}, Gaussian noise) into samples from the target data distribution. Formally, given data samples $\boldsymbol{x}_1 \sim p_{\text{data}}$ and random noise samples $\boldsymbol{x}_0 \sim p_{\text{prior}}$, Flow Matching trains a time-dependent vector field $\mathcal{V}_{\boldsymbol{\theta}}(\boldsymbol{x}_t, t)$ to approximate the optimal transport velocity between $\boldsymbol{x}_0$ and $\boldsymbol{x}_1$. The intermediate samples $\mathbf{x}_t$ are defined as a linear interpolation:
\begin{equation}
    \begin{split}
        \boldsymbol{x}_t = (1-t)\boldsymbol{x}_0 + t\boldsymbol{x}_1,
    \end{split}
\end{equation}
where $t\in [0,1]$ denotes the time step.
The objective is to minimize the squared error between the predicted velocity and the true velocity derived from the interpolation path:
\begin{equation}
    \begin{split}
        \mathcal{L}_{\text{FM}} = \mathbb{E}_{t,\boldsymbol{x}_0,\boldsymbol{x}_1}\left[ ||\mathcal{V}_{\boldsymbol{\theta}}(\boldsymbol{x}_t, t) - (\boldsymbol{x}_1 - \boldsymbol{x}_0)||_2^2 \right].
    \end{split}
\end{equation}
During inference, the model generates samples by solving the following ordinary differential equation (ODE) in time:
\begin{equation}
    \begin{split}
        \frac{d\boldsymbol{x}_t}{dt} = \mathcal{V}_{\boldsymbol{\theta}}(\boldsymbol{x}_t, t),
    \end{split}
\end{equation}
starting from a noise sample $\boldsymbol{x}_0 \sim p_{\text{prior}}$ and integrating from $t=0$ to $t=1$ to obtain the generated data sample $\hat{\boldsymbol{x}}_1$. In practice, the integration can be efficiently implemented using standard numerical solvers such as Euler’s method.

\noindent
\textbf{Audio encoding and decoding.}
Following previous works~\cite{kushwaha2025vintage,cheng2025mmaudio,shan2025hunyuanvideo,liu2023audioldm,lee2024voiceldm}, we perform the flow matching process in the audio latent space to enhance training stability and generation efficiency.
Specifically, the raw audio waveforms are first converted into Mel-spectrograms $\boldsymbol{m}$ using the Short-Time Fourier Transform (STFT). Then, these Mel-spectrograms are encoded into latent representations using a pretrained VAE~\cite{kingma2013auto}, which compresses the high-dimensional spectrograms into a lower-dimensional latent space. We adopt the pre-trained Audio VAE from AudioLDM~\cite{liu2023audioldm}.
This process can be denoted as $\boldsymbol{x}=\mathcal{E}(\boldsymbol{m})$, where $\mathcal{E}$ is the encoder of the VAE.
During inference, the denoised latent representation $\hat{\boldsymbol{x}}$ is passed through the VAE decoder to reconstruct the corresponding Mel-spectrogram $\hat{\boldsymbol{m}} = \mathcal{D}(\hat{\boldsymbol{x}})$, which is then converted into a time-domain waveform using a vocoder (\textit{e.g.}, HiFi-GAN~\cite{kong2020hifi}).

\begin{figure*}
    \centering
    \includegraphics[width=0.98\textwidth]{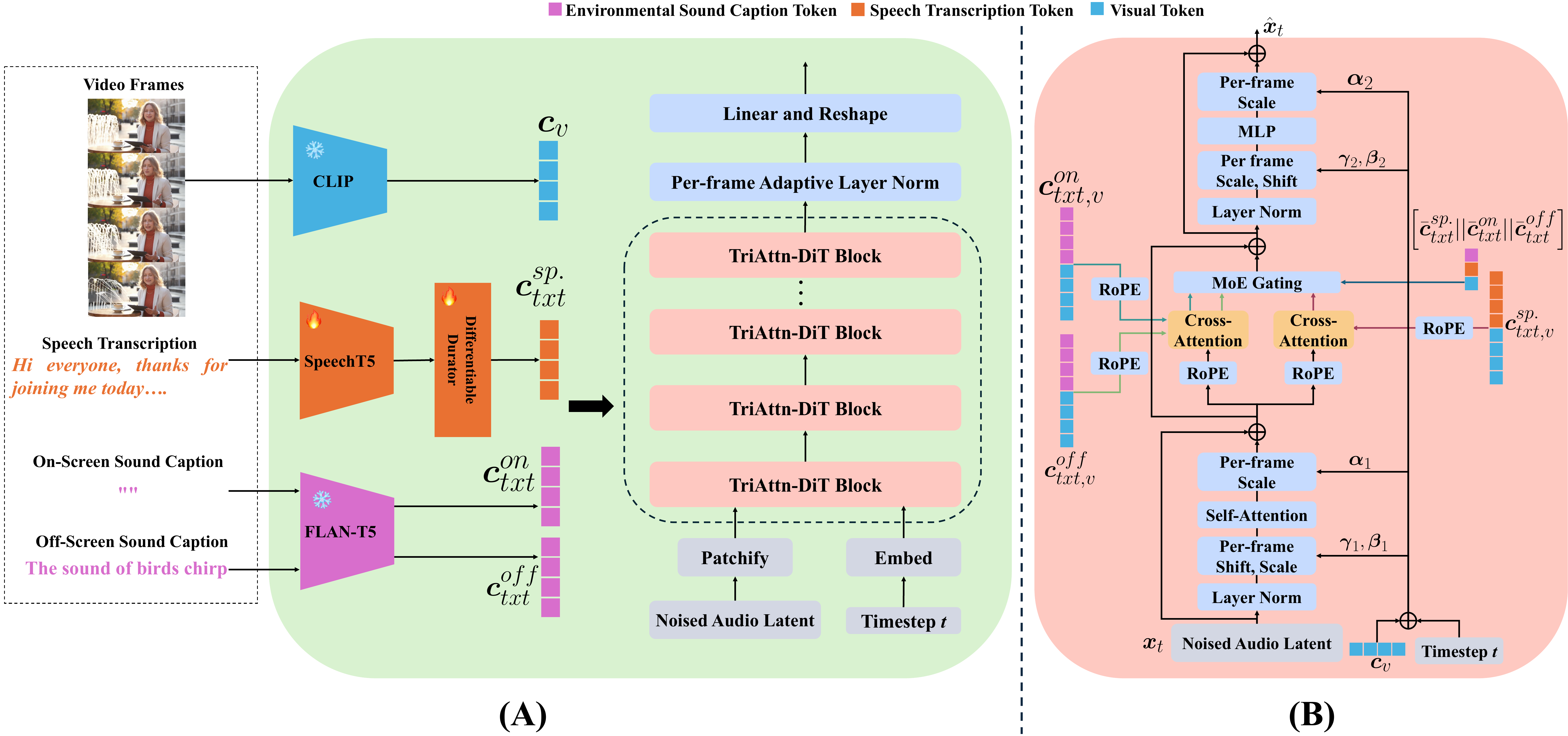}
    \vspace{-9pt}
    \caption{(A) Overview of our proposed OmniSonic, which mainly consists of an environmental text encoder (FLAN-T5), a speech transcription encoder (SpeechT5), a visual encoder (CLIP visual encoder), an audio VAE, and our specially designed TriAttn-DiT blocks. The input example demonstrates the scenario of on-screen speech with off-screen environmental sound. The input conditions include visual frames, speech transcription, an on-screen environmental sound caption (represented by a placeholder \texttt{""}), and an off-screen environmental sound caption.
    (B) Details of our proposed TriAttn-DiT block.}
    \label{fig:overview}
    \vspace{-12pt}
\end{figure*}

\subsection{Overview}

The overview of the OmniSonic is illustrated in Fig.~\ref{fig:overview}.
In this subsection, we introduce each component and describe the overall computation process.

\subsubsection{Condition Encoders}


The condition encoders in OmniSonic aim to encode both text and visual inputs into embeddings that serve as conditioning signals for the diffusion backbone. Given an on-screen environmental sound caption $\boldsymbol{txt}_{on}^{env.}$, an off-screen environmental sound caption $\boldsymbol{txt}_{off}^{env.}$, a speech transcription $\boldsymbol{txt}^{sp.}$, and visual frames $\boldsymbol{f}$, the encoders take these inputs and produce their corresponding representations:
\begin{equation}
    \begin{split}
        \boldsymbol{c}_{txt}^{on} = E_{env.}(\boldsymbol{txt}_{on}^{env.}), \quad &\boldsymbol{c}_{txt}^{off} = E_{env.}(\boldsymbol{txt}_{off}^{env.}), \\
        \boldsymbol{c}_{txt}^{sp} = E_{sp.}(\boldsymbol{txt}^{sp.}), &\quad \boldsymbol{c}_v = E_v(\boldsymbol{f}),
    \end{split}
\end{equation}
where $\boldsymbol{c}_{txt}^{on}\in \mathbb{R}^{L_{on}\times d_{env.}}$, $\boldsymbol{c}_{txt}^{off} \in \mathbb{R}^{L_{off}\times d_{env.}}$, $\boldsymbol{c}_{txt}^{sp.} \in \mathbb{R}^{L_{sp.}\times d_{sp.}}$, and $\boldsymbol{c}_v\in \mathbb{R}^{N\times d_{vis.}}$ denote the embeddings of their corresponding input conditions and $E_{env}$, $E_{sp.}$, and $E_{v}$ represent the encoders for environmental sound caption, speech transcription, and visual frames, respectively.
The environmental encoder captures text descriptions of sound events, the speech encoder models linguistic content from transcriptions, and the visual encoder provides temporal and spatial cues from the video. 
Together, they enable OmniSonic to integrate heterogeneous modalities for universal holistic audio generation.

In implementation, we adopt FLAN-T5~\cite{raffel2020exploring} as the environmental sound caption encoder and the CLIP visual encoder~\cite{radford2021learning} for obtaining visual frame features. For the speech transcription encoder, we follow the setup in VoiceLDM~\cite{lee2024voiceldm} and employ SpeechT5~\cite{ao2022speecht5}, followed by a differentiable Durator module~\cite{lee2024voiceldm,tan2024naturalspeech} to upsample the hidden sequence. During training, the environmental sound caption encoder and the visual encoder are frozen, while the speech transcription encoder and the differentiable durator are trainable following~\cite{lee2024voiceldm}. A simple illustration is presented in left part of Fig.~\ref{fig:overview}, where the illustrated input example demonstrates the scenario of on-screen speech with off-screen environmental sound, the condition of which includes visual frames, speech transcription, an on-screen environmental sound caption (represented by a placeholder \texttt{""}), and an off-screen environmental sound caption.

\subsubsection{TriAttn-DiT}

The TriAttn-DiT serves as the diffusion backbone of our OmniSonic framework. It integrates information from visual, on/off-screen environmental text, and speech transcriptions through cross-attentions with MoE gating to predict the audio latent velocity for audio generation.

\vspace{1mm}
\noindent
\textbf{Frame-Aligned Adaptive Layer Normalization.}
Condition embeddings are fed to the TriAttn-DiT backbone, which stacks multiple TriAttn-DiT blocks. In each block, the noisy audio latent representation $\boldsymbol{x}_t$ is first processed by an adaptive layer normalization (adaLN) layer~\cite{peebles2023scalable}. Inspired by the frame-aligned adaLN mechanism~\cite{cheng2025mmaudio}, which enhances audio–visual synchronization, we adopt a similar strategy to further improve the temporal alignment between the generated audio and the conditioned video frames.

Specifically, given the visual condition $\boldsymbol{c}_v \in \mathbb{R}^{N \times d_{\text{vis}}}$ and the time step embedding $\boldsymbol{t} \in \mathbb{R}^d$, we project the visual condition into the same dimensional space as the time embedding and then perform element-wise addition: $\boldsymbol{c}_{vt} = \boldsymbol{c}_v \boldsymbol{W}_{vt} + \boldsymbol{t}$,
where $\boldsymbol{W}_{vt} \in \mathbb{R}^{d_{\text{vis}} \times d}$ denotes the learnable projection weights, $\boldsymbol{c}_{vt}\in \mathbb{R}^{N\times d}$ is the resulting fused representation. After that, $\boldsymbol{c}_{vt}$ is fed into the adaLN layer, in which the nearest-neighbor interpolation is applied to to upsample the frame rate of $\boldsymbol{c}_{vt}$, aligning it with the temporal resolution of the audio latent $\boldsymbol{x}_t$:
\begin{equation}
    \begin{split}
        [\boldsymbol{\alpha}_1, \boldsymbol{\beta}_1, \boldsymbol{\gamma}_1, \boldsymbol{\alpha}_2, \boldsymbol{\beta}_2, \boldsymbol{\gamma}_2] = \text{Upsample}(\text{Proj}_{d\rightarrow 6d}(\boldsymbol{c}_{vt})),
    \end{split}
\end{equation}
where $\boldsymbol{\alpha}_1$, $\boldsymbol{\beta}_1$, $\boldsymbol{\gamma}_1$, $\boldsymbol{\alpha}_2$, $\boldsymbol{\beta}_2$, $\boldsymbol{\gamma}_2$ are generated per-frame adaLN parameters.
These parameters will modulate subsequent layers within the TriAttn-DiT block.

\vspace{1mm}
\noindent
\textbf{TriCrossAttention.}
The TriCrossAttention module is the core component of the TriAttn-DiT block, responsible for integrating multi-condition information from on-screen and off-screen environmental captions as well as speech transcriptions. It dynamically computes cross-modal interactions based on the provided conditioning content, enabling effective fusion and alignment across different auditory modalities. 
Given the conditions $\boldsymbol{c}_{txt}^{on}$, $\boldsymbol{c}_{txt}^{off}$, $\boldsymbol{c}_{txt}^{sp.}$, and $\boldsymbol{c}_v$, the visual condition $\boldsymbol{c}_{txt}^{sp.}$ is concatenated with the corresponding on-screen conditions. Specifically, when the provided original on-screen environmental sound caption text is not empty (\textit{i.e.}, not \texttt{""}), indicating that the visual information is associated with the on-screen environmental sound, we perform the following operation:
\begin{equation}
    \begin{split}
        \boldsymbol{c}^{on}_{txt,v} = [\boldsymbol{c}^{on}_{txt}|| \boldsymbol{c}_v & \boldsymbol{W}_{env.}], \quad \boldsymbol{c}^{off}_{txt,v} = [\boldsymbol{c}^{off}_{txt}|| \boldsymbol{0}\boldsymbol{W}_{env.}], \\
        &\boldsymbol{c}^{sp.}_{txt,v} =  [\boldsymbol{c}^{sp.}_{txt}|| \boldsymbol{0}\boldsymbol{W}_{sp.}], \\
    \end{split}
    \label{eq:text_vis_concat}
\end{equation}
where $\boldsymbol{c}^{on}_{txt,v}\in \mathbb{R}^{(L_{on}+N)\times d_{env.}}$, $\boldsymbol{c}^{off}_{txt,v} \in \mathbb{R}^{(L_{off}+N)\times d_{env.}}$, and $\boldsymbol{c}^{sp.}_{txt,v} \in \mathbb{R}^{(L_{sp.}+N)\times d_{sp.}}$ represent the concatenated embeddings for on-screen, off-screen, and speech conditions, respectively. $\boldsymbol{W}_{env.}\in \mathbb{R}^{d_v\times d_{env.}}$ and $\boldsymbol{W}_{sp.}\in \mathbb{R}^{d_v\times d_{sp.}}$ denote learnable projection matrices. $\boldsymbol{0}\in \mathbb{R}^{N\times d_{vis.}}$ is an all-zero matrix used as a placeholder when the visual feature is not associated with the corresponding condition.
In contrast, when the provided on-screen environmental sound caption text is empty (\textit{i.e.}, \texttt{""}), indicating that the visual information corresponds to the speech rather than an environmental sound. Then $\boldsymbol{c}^{on}_{txt,v}$ and $\boldsymbol{c}^{sp.}_{txt,v}$ in Eq.~\ref{eq:text_vis_concat} become $[\boldsymbol{c}^{on}_{txt}|| \boldsymbol{0}\boldsymbol{W}_{env.}]$ and $[\boldsymbol{c}^{sp.}_{txt}|| \boldsymbol{c}_v\boldsymbol{W}_{sp.}]$, respectively, where the visual feature $\boldsymbol{c}_v$ is fused with the speech condition to enhance temporal and semantic alignment between the visible speaking person and the generated speech audio.

After obtaining the visual-text concatenated embeddings, we conduct the cross-attention operations between them and the audio latent representation $\boldsymbol{x}_t$. To incorporate positional information and enhance temporal correspondence, we first apply the Rotary Position Embedding (RoPE)~\cite{su2024roformer} to both $\boldsymbol{x}_t$ and the visual tokens within the text–visual concatenated embeddings $\boldsymbol{c}^{on}_{txt,v}$, $\boldsymbol{c}^{off}_{txt,v}$, and $\boldsymbol{c}^{sp.}_{txt,v}$, followed by triple cross-attention operations:
\begin{equation}
    \begin{split}
        \boldsymbol{x}_t^{on}= \text{CA}_{env.}(\text{RoPE}(\boldsymbol{x}_t),  &\text{RoPE}(\boldsymbol{c}^{on}_{txt,v}[L_{on}:,  :]), \boldsymbol{c}^{on}_{txt,v}), \\
        \boldsymbol{x}_t^{off} = \text{CA}_{env.}(\text{RoPE}(\boldsymbol{x}_t), &\text{RoPE}(\boldsymbol{c}^{off}_{txt,v}[L_{off}:, :]), \boldsymbol{c}^{off}_{txt,v}), \\
        \boldsymbol{x}_t^{sp.} = \text{CA}_{sp.}(\text{RoPE}(\boldsymbol{x}_t),  &\text{RoPE}(\boldsymbol{c}^{sp.}_{txt,v}[L_{sp.}:, :]), \boldsymbol{c}^{sp.}_{txt,v}),
    \end{split}
\end{equation}
where $\text{CA}(Q,K,V)$ denotes the cross-attention operation, and $Q$, $K$, and $V$ are the query, key, and value matrices, respectively. Here $\boldsymbol{x}_t$ serves as the query, and the full concatenated embeddings $\boldsymbol{c}^{on}_{txt,v}$, $\boldsymbol{c}^{off}_{txt,v}$, and $\boldsymbol{c}^{sp.}_{txt,v}$ are used as both keys and values. Importantly, RoPE is only applied to the visual token segments, \textit{i.e.}, $\boldsymbol{c}^{on}_{txt,v}[L_{on}:, :]$, $\boldsymbol{c}^{off}_{txt,v}[L_{off}:, :]$, and $\boldsymbol{c}^{sp.}_{txt,v}[L_{sp.}:, :]$, to encode positional and temporal information, while the textual part remains unchanged. This allows the model to retain semantic consistency from textual tokens while incorporating temporal alignment cues from visual tokens during the attention computation.

\vspace{1mm}
\noindent
\textbf{MoE Gating.} To effectively fuse the attended audio latent representations $\boldsymbol{x}_t^{on}$, $\boldsymbol{x}_t^{off}$, and $\boldsymbol{x}_t^{sp.}$, we incorporate a Mixture-of-Experts (MoE)-based Gating module in each TriAttn-DiT block. This module adaptively computes a weighted sum of the three representations based on their corresponding condition embeddings, enabling dynamic balancing among different condition types during generation. Specifically, given the condition embeddings $\boldsymbol{c}^{on}_t$, $\boldsymbol{c}^{off}_t$, and $\boldsymbol{c}^{sp.}_t$, we first compute the mean token by averaging along the sequence length dimension, yielding a single representative token for each condition. The three resulting tokens are then concatenated and passed through a lightweight MLP, followed by a softmax function to obtain three normalized gating weights:
\begin{equation}
    \begin{split}
        [\omega^{sp.}, \omega^{on}, \omega^{off}] = \text{Softmax}(\text{MLP}([\bar{\boldsymbol{c}}^{sp.}_{txt} || \bar{\boldsymbol{c}}^{on}_{txt} || \bar{\boldsymbol{c}}^{off}_{txt}])),
    \end{split}
\end{equation}
where $\bar{\boldsymbol{c}}^{sp.}_{txt}$, $\bar{\boldsymbol{c}}^{on}_{txt}$, and $\bar{\boldsymbol{c}}^{off}_{txt}$ denote the mean tokens of $\boldsymbol{c}^{sp.}_{txt}$, $\boldsymbol{c}^{on}_{txt}$, and $\boldsymbol{c}^{off}_{txt}$, respectively.
The resulting gating weights $\omega^{sp.}$, $\omega^{on}$, and $\omega^{off}$ are then used to adaptively fuse the three attended latent representations through a weighted summation: 
$\boldsymbol{v}_t = \omega^{sp.}\boldsymbol{x}^{sp.}_t + \omega^{on}\boldsymbol{x}^{on}_t + \omega^{off}\boldsymbol{x}^{off}_t$, where $\boldsymbol{v}_t$ denote the predicted velocity at time step $t$, which is used by the ODE-based sampling procedure during inference to update the audio latent representation.
This MoE gating mechanism allows each TriAttn-DiT block to dynamically adjust its focus among different textual condition types according to the contextual relevance at each time step, ensuring coherent and balanced conditioning during audio generation.

After obtaining the denoised latent audio representation $\hat{\boldsymbol{x}}_1$, we use the VAE decoder to reconstruct the Mel-spectrogram, $\hat{\boldsymbol{m}} = \mathcal{D}(\hat{\boldsymbol{x}}_1)$, followed by a HiFi-GAN~\cite{kong2020hifi} vocoder to convert it into a time-domain waveform.

\section{Experiments}

\subsection{Experimental Setup}

\begin{table*}[htbp]
  \centering
  \caption{Experimental results of our OmniSonic and compared baselines on the proposed UniHAGen-Bench using objective evaluation metrics. Bold values denote the best performance in each column.}
  \vspace{-8pt}
  \resizebox{\textwidth}{!}{
    \begin{tabular}{lccccccccccc}
    \toprule
    \multirow[c]{2}{*}{Model}
      & \multicolumn{2}{c}{Condition Modality}
      & \multicolumn{2}{c}{Generation Quality}
      & \multicolumn{3}{c}{Semantic Alignment}
      & \multicolumn{3}{c}{Speech Correctness}
      & \multicolumn{1}{c}{Temporal Alignment} \\
    \cmidrule(lr){2-3}\cmidrule(lr){4-5}\cmidrule(lr){6-8}\cmidrule(lr){9-11}\cmidrule(lr){12-12}
    \multicolumn{1}{c}{} 
      & \multicolumn{1}{c}{Text} & \multicolumn{1}{c}{Visual}
      & \multicolumn{1}{c}{FAD$\downarrow$} & \multicolumn{1}{c}{MKL$\downarrow$}
      & \multicolumn{1}{c}{AT score$\uparrow$} & \multicolumn{1}{c}{AV score$\uparrow$} & \multicolumn{1}{c}{Mean$\uparrow$}
      & \multicolumn{1}{c}{WER$\downarrow$} & \multicolumn{1}{c}{CER$\downarrow$} & \multicolumn{1}{c}{PER$\downarrow$}
      & \multicolumn{1}{c}{DeSync$\downarrow$} \\
    \midrule
    AudioLDM 2~\cite{liu2024audioldm} & \Checkmark     & \XSolidBrush     & 10.28 & 6.37  & 20.17 & 6.62  & 13.40  & 1.00   & 1.01  & 0.97  & 1.39 \\
    AudioLDM 2-gigaspeech~\cite{liu2024audioldm} & \Checkmark     & \XSolidBrush     & 14.09 & 9.33  & 24.26 & 7.65  & 15.96 & 0.66  & 0.50   & 0.51  & 1.04 \\
    VoiceLDM~\cite{lee2024voiceldm} & \Checkmark     & \XSolidBrush     & 3.58  & 5.74  & 21.25 & 6.80   & 14.03 & 0.15  & 0.08  & 0.09  & 1.25 \\
    VinTAGe~\cite{kushwaha2025vintage} & \Checkmark     & \Checkmark     & 8.30   & 5.00     & 20.60  & 8.36  & 14.48 & 1.99  & 1.46  & 1.36  & 1.35 \\
    MMAudio~\cite{cheng2025mmaudio} & \Checkmark     & \Checkmark     & 5.82  & 5.60   & 23.37 & \textbf{11.12} & 17.25 & 1.50   & 1.31  & 1.19  & 0.51 \\
    HunyuanVideo-Foley~\cite{shan2025hunyuanvideo} &   \Checkmark    & \Checkmark     & 6.00     & 5.88  & 23.03 & 10.86 & 16.95 & 1.36  & 1.13  & 1.13  & \textbf{0.38} \\
    \midrule
    \textbf{OmniSonic (Ours)} &   \Checkmark   & \Checkmark    & \textbf{3.07}  & \textbf{2.79}  & \textbf{28.07} & 9.01  & \textbf{18.54} & \textbf{0.14}  & \textbf{0.05}  & \textbf{0.06}  & 0.72 \\
    \bottomrule
    \end{tabular}%
    }
    \vspace{-15pt}
  \label{tab:main_res}%
\end{table*}%

\begin{table}[htbp]
  \centering
  \caption{Experimental results of our OmniSonic and compared baselines using objective evaluation metrics. Bold values denote the best performance in each column.}
  \vspace{-8pt}
  \resizebox{0.48\textwidth}{!}{
    \begin{tabular}{lcccc}
    \toprule
    Model & MOS-Q$\uparrow$ & MOS-EF$\uparrow$ & MOS-SF$\uparrow$ & MOS-T$\uparrow$ \\
    \midrule
    VoiceLDM~\cite{lee2024voiceldm} & 3.13  & 3.40   & 4.05  & 2.54 \\
    MMAudio~\cite{cheng2025mmaudio} & 3.74  & 3.24  & 1.15  & 3.71 \\
    HunyuanVideo-Foley~\cite{shan2025hunyuanvideo} & 3.85  & 3.88  & 1.17  & 3.86 \\
    \midrule
    \textbf{OmniSonic (Ours)} & \textbf{4.35}  & \textbf{4.42}  & \textbf{4.74}  & \textbf{4.29} \\
    \bottomrule
    \end{tabular}%
    }
    \vspace{-18pt}
  \label{tab:user_study}%
\end{table}%

\textbf{Dataset and Benchmark.}
Since no existing dataset that fully aligns with the scenarios defined in \textbf{UniHAGen}, we construct our training data by synthesizing samples from publicly available datasets. Specifically, we utilize VGGSound~\cite{chen2020vggsound} as the audio–visual environmental sound source, and LRS3~\cite{afouras2018lrs3} as the audio–visual speech source. To enrich the diversity of speech content and enhance the model’s generalization ability in speech-related scenarios, we incorporate CommonVoice~\cite{ardila2020common} as an additional text-to-speech corpous.
For VGGSound, we exclude all speech-related categories and remove samples whose captions contain speech-related keywords (\textit{e.g.},``speech", ``voice", ``say", etc.), ensuring that the remaining clips primarily represent non-speech environmental sounds. The resulting dataset consists of approximately 195K, 33K, and 1.67M samples for VGGSound, LRS3, and CommonVoice, respectively.

To simulate the three scenarios defined in UniHAGen, we combine clips from these datasets. 
For Scenario 1, we sample a video clip with environmental sound as the on-screen content and mix it with a randomly selected human speech clip as the off-screen sound. For Scenario 2, we select a speaking-person video from LRS3 as the on-screen speech component and mix it with a randomly sampled environmental sound from VGGSound. For Scenario 3, we mix an on-screen environmental video with both an additional off-screen environmental sound and an off-screen speech clip, forming complex multi-source scenes. All mixtures are performed at randomly sampled signal-to-noise ratio (SNR) levels to increase acoustic diversity.
Finally, each training sample consists of the mixed waveform, corresponding textual conditions (environmental captions and speech transcriptions), and visual frames. This synthetic construction provides diverse and well-aligned examples that closely match the requirements of the UniHAGen task.

To comprehensively evaluate model performance under our UniHAGen setting, we construct a benchmark named UniHAGen-Bench. The samples are collected from the testing splits of VGGSound and LRS3, ensuring that none of them overlap with the training data. The benchmark consists of 1,003 manually curated samples covering the three representative scenarios in our task: (1) Scenario 1: on-screen environmental sound + off-screen human speech (300 samples), (2) Scenario 2: on-screen human speech + off-screen environmental sound (401 samples), and (3) Scenario 3: on-screen environmental sound + off-screen environmental sound + off-screen human speech (302 samples). Each sample contains synchronized video frames, textual conditions (environmental captions and speech transcriptions), and mixed audio constructed under controlled signal-to-noise ratio (SNR) levels. This benchmark establishes a standardized and diverse evaluation foundation for assessing the fidelity and cross-modal consistency of audio generation models under the UniHAGen setting.


\vspace{1mm}
\noindent
\textbf{Evaluation Metrics.} We evaluate OmniSonic and the baselines using the objective evaluation metrics of Fr{\'e}chet Audio Distance (FAD)~\cite{kilgour2018fr}, Mean Kullback–Leibler Divergence (MKL)~\cite{iashin2021taming}, AV score~\cite{kushwaha2025vintage}, AT score~\cite{kushwaha2025vintage}, Word Error Rate~\cite{lee2024voiceldm}, Character Error Rate (CER)~\cite{kim2022guided}, Phoneme Error Rate (PER)~\cite{lee2022hierspeech}, and DeSync score~\cite{cheng2025mmaudio,shan2025hunyuanvideo}, and subjective evaluation metrics of overall quality (MOS-Q), environmental faithfulness (MOS-EF), speech faithfulness (MOS-SF), and temporal alignment (MOS-T). Details of these metrics and the subjective evaluation process are provided in the Appendix.

\vspace{1mm}
\noindent
\textbf{Baselines.}
We compare OmniSonic with SOTA audio generation models: AudioLDM 2~\cite{liu2024audioldm}, VoiceLDM~\cite{lee2024voiceldm}, VinTAGe~\cite{kushwaha2025vintage}, MMAudio~\cite{cheng2025mmaudio}, and HunyuanVideo-Foley~\cite{shan2025hunyuanvideo}. Details are provided in the Appendix.

\begin{figure*}[htbp]
    \centering
    \includegraphics[width=0.77\textwidth]{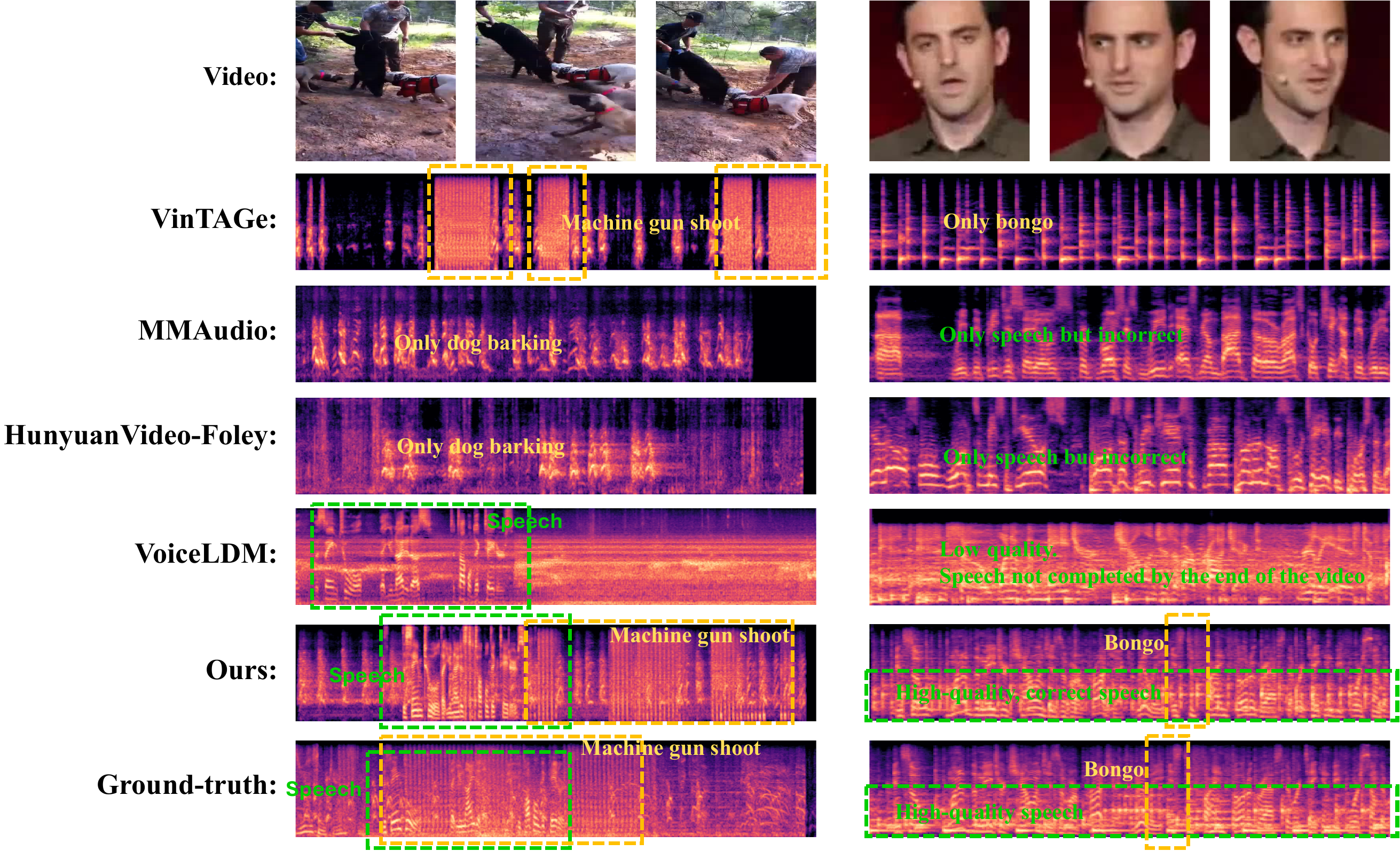}
    \vspace{-5pt}
    \caption{Visualization of the spectrograms of generated audios and the ground-truth.}
    \label{fig:vis}
    \vspace{-13pt}
\end{figure*}

\subsection{Experimental Results and Analysis}

\paragraph{Objective Evaluation.}
Objective evaluation results on UniHAGen-Bench for our proposed OmniSonic and the compared baselines are presented in Tab.~\ref{tab:main_res}. OmniSonic achieves state-of-the-art performance across nearly all metrics. For audio quality, OmniSonic surpasses the best baseline by \textbf{0.51} and \textbf{2.21} in terms of FAD and MKL, respectively. For semantic alignment, the mean of AT and AV scores exceeds MMAudio by \textbf{1.29}. In terms of speech correctness, our model also outperforms VoiceLDM across all three speech-related metrics (WER, CER, and PER). 

\noindent
\underline{\textit{Limitations of Baselines.}}
We further analyze the performance gaps among different baselines. VinTAGe, MMAudio, and HunyuanVideo-Foley mainly focus on environmental sound synthesis, lacking explicit modeling of human speech; hence, they fail on speech-related metrics and perform poorly in mixed speech–environment scenarios. VoiceLDM, though effective at speech generation, relies solely on textual input without visual grounding, resulting in low-quality background sound and poor on/off-screen distinction. AudioLDM 2, designed for general text-to-audio generation, struggles to capture fine-grained audio–visual correspondence, leading to weaker AT and AV scores. In contrast, OmniSonic benefits from its joint text–video conditioning, TriAttn-DiT architecture, and MoE-based Gating mechanism, which collaboratively enable adaptive fusion and dynamic weighting of multiple conditioning signals from both on-screen and off-screen sources. It also allows our model to coherently model both speech and environmental sounds, achieving universal holistic audio generation that captures complex real-world auditory interactions within a unified diffusion framework.

\noindent
\underline{\textit{Analysis on Temporal Alignment Score.}}
For the temporal alignment score (DeSync), our OmniSonic does not achieve the best result compared with MMAudio and HunyuanVideo-Foley. This difference can be attributed to the utilization of visual features. Specifically, in addition to CLIP-based visual embeddings, MMAudio and HunyuanVideo-Foley also incorporate visual features extracted from Synchformer~\cite{iashin2024synchformer}, which provide fine-grained temporal cues for audio–visual synchronization. In contrast, our model relies solely on CLIP visual features for visual conditioning, which excel at capturing global semantic context but are less effective at modeling detailed local and spatial correspondences. Notably, CLIP-based visual features tend to remain almost invariant across consecutive frames when the on-screen objects persist without noticeable visual changes. While this makes them effective at capturing the overall scene semantics and transitions, such as when a new sound-emitting object appears on screen, they provide limited temporal sensitivity in visually stable or weakly dynamic scenes, leading to less precise audio–visual synchronization in such cases. Moreover, since the DeSync metric computes temporal alignment using Synchformer features, models that incorporate these features (\textit{e.g.}, MMAudio and HunyuanVideo-Foley) may benefit slightly from a closer feature distribution, potentially yielding higher scores under this metric.

\vspace{1mm}
\noindent
\textbf{Subjective Evaluation.}
The results of subjective evaluation are presented in Tab.~\ref{tab:user_study}. Our OmniSonic consistently achieves the highest scores across all aspects, including overall quality (MOS-Q), environmental faithfulness (MOS-EF), speech faithfulness (MOS-SF), and temporal alignment (MOS-T). Compared with the baselines, OmniSonic produces audio that is perceptually more natural and semantically better aligned with both on-screen and off-screen events. In particular, the substantial improvements in MOS-SF and MOS-EF demonstrate that our model effectively handles mixed scenarios involving speech and environmental sounds, generating coherent and realistic auditory scenes across diverse conditions.

\subsection{Qualitative Analysis}

We present qualitative results in Fig.~\ref{fig:vis}. The left example depicts a scene with the on-screen environmental sound caption \textit{``Dogs bark and growl as they surround a bluetick, creating a chaotic and intense atmosphere in an archaeological excavation site''}, the the off-screen environmental sound caption \textit{``The sound of machine gun shooting''}, and an accompanying off-screen speech. The right example shows a case with on-screen speech and an off-screen environmental sound caption \textit{``The sound of a rhythmic beat of bongo''}.

\begin{table*}[t]
  \centering
  \caption{Ablation study on the MoE Gating. Bold values denote the best performance in each column.}
  \vspace{-8pt}
  \resizebox{\textwidth}{!}{
    \begin{tabular}{lccccccccc}
    \toprule
    \multirow[c]{2}{*}{Model}
      & \multicolumn{2}{c}{Generation Quality}
      & \multicolumn{3}{c}{Semantic Alignment}
      & \multicolumn{3}{c}{Speech Correctness}
      & \multicolumn{1}{c}{Temporal Alignment} \\
    \cmidrule(lr){2-3}\cmidrule(lr){4-6}\cmidrule(lr){7-9}\cmidrule(lr){10-10}
    \multicolumn{1}{c}{} 
      & \multicolumn{1}{c}{FAD$\downarrow$} & \multicolumn{1}{c}{MKL$\downarrow$}
      & \multicolumn{1}{c}{AT score$\uparrow$} & \multicolumn{1}{c}{AV score$\uparrow$} & \multicolumn{1}{c}{Mean$\uparrow$}
      & \multicolumn{1}{c}{WER$\downarrow$} & \multicolumn{1}{c}{CER$\downarrow$} & \multicolumn{1}{c}{PER$\downarrow$}
      & \multicolumn{1}{c}{DeSync$\downarrow$} \\
    \midrule
    \textbf{OmniSonic} & \textbf{3.07}  & \textbf{2.79}  & \textbf{28.07} & \textbf{9.01}  & \textbf{18.54} & \textbf{0.14}  & \textbf{0.05}  & \textbf{0.06}  & \textbf{0.72} \\
    \textit{w/o MoE Gating} & 6.12  & 5.25  & 24.76 & 7.11  & 15.94 & 0.56  & 0.49  & 0.49  & 1.23 \\
    \bottomrule
    \end{tabular}%
    }
    \vspace{-12pt}
  \label{tab:ablation}%
\end{table*}%

For the left example, we observe that VinTAGe can generate the off-screen environmental sound but fails to synthesize speech. MMAudio and HunyuanVideo-Foley focus primarily on the on-screen sound, neglecting off-screen auditory events. VoiceLDM is able to synthesize speech in this case, however, the generated speech is of low quality, and the environmental sound resembles artificial noise or artifacts rather than realistic ambient audio. 
In contrast, the audio generated by OmniSonic successfully synthesizes all sound components, including the on-screen environmental sound, the off-screen environmental sound, and the off-screen speech, with high perceptual quality and natural blending. The speech is content-accurate with respect to the transcription, while the environmental sounds faithfully reflect both on- and off-screen events without introducing noticeable artifacts.
For the right example, while VinTAGe can handle off-screen events, it still fails to generate any speech. MMAudio and HunyuanVideo-Foley do not produce off-screen sounds, and their generated speech does not match the provided transcription. VoiceLDM is able to generate both speech and environmental sound, however, the overall audio quality is low, and the generated speech is incomplete before the video ends. 
In contrast, OmniSonic successfully generates both the on-screen speech and the off-screen environmental sound with high perceptual quality, and the generated speech is complete and accurate according to the provided transcription.

These results show OmniSonic’s capability to produce high-quality and complete auditory scenes across diverse sound domains and complex multi-source conditions.

\begin{figure}[t]
    \centering
    \includegraphics[width=0.48\textwidth]{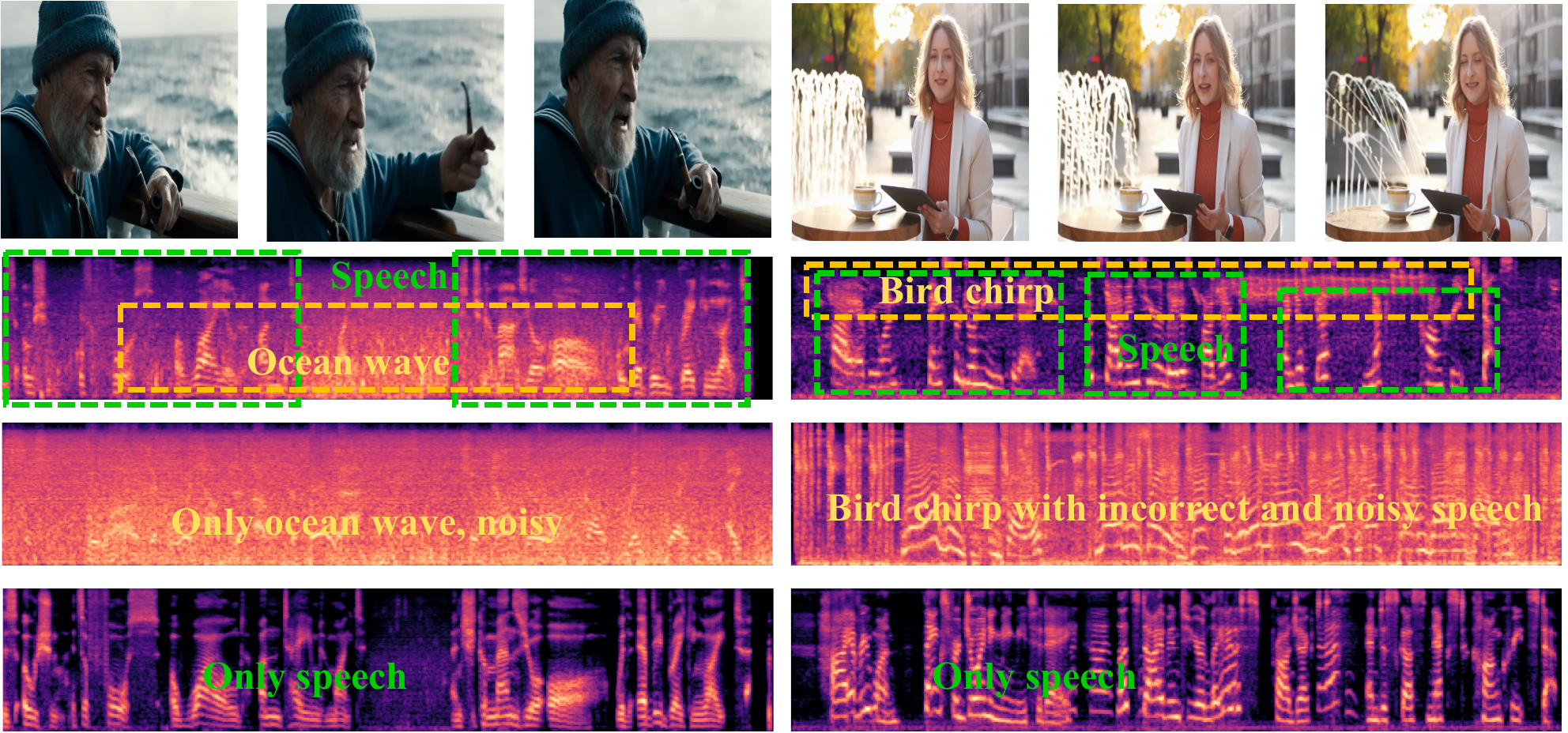}
    \vspace{-18pt}
    \caption{Ablation study on the MoE Gating module using in-the-wild samples. In this example, sounds are generated with different gating configurations. Top row: MoE Gating (full model); Middle row: reduced weight for the speech branch; Bottom row: reduced weight for the off-screen environmental branch.}
    \label{fig:ablation_real}
    \vspace{-4.6mm}
\end{figure}
\subsection{Ablation Study}

We further conduct an ablation study on the Mixture-of-Experts (MoE) Gating module within the TriAttn-DiT blocks to evaluate its impact on model performance and the adaptive fusion of multi-source conditions. As shown in Tab.~\ref{tab:ablation}, removing the MoE Gating module leads to substantial performance degradation across all metrics. Specifically, generation quality declines sharply (FAD increases from 3.07 to 6.12, MKL from 2.79 to 5.25), and semantic alignment also drops notably, with the mean of AT and AV scores decreasing from 18.54 to 15.94. In terms of speech correctness, the error rates (WER, CER, and PER) increase by more than 3×, indicating that the gating mechanism plays a crucial role in balancing the influence of speech- and environment-related conditions. Temporal alignment (DeSync) also worsens without MoE Gating, suggesting that dynamic weighting among conditional branches helps the model maintain temporal coherence between generated audio and visual context. These results demonstrate that the MoE Gating module is essential for achieving coherent, high-fidelity, and semantically aligned audio generation across diverse on- and off-screen conditions.

To further investigate the effect of the MoE Gating module, we manually set the scores of different conditions generated by the MLP layer (before the Softmax) to small values, \textit{e.g.} 0.0, to explore cases where the gating mechanism provides limited contributions from certain modalities. 
We perform this analysis using in-the-wild samples, rather than the synthesized benchmark examples constructed from VGGSound and LRS3. We show the results in Fig.~\ref{fig:ablation_real}, in which the left sample is with the on-screen speech transcription of \textit{``This ocean, it's a force, a wild, untamed might. And she commands your awe with every breaking light"} and off-screen environmental sound caption of \textit{``The sea breeze blows on the sea, raising waves"}, while the right sample is with the on-screen speech transcription of \textit{``Hi everyone, thanks for joining me today, let's dive into your latest project and discuss next concrete steps"} and off-screen environmental sound caption of \textit{``The sound of birds chirp"}.
As illustrated, when the speech branch is suppressed (second row), the model fails to generate meaningful speech or only produces noisy environmental sound. Conversely, when the off-screen environmental branch is suppressed (third row), the model can generate speech but loses the off-screen environmental sound component, resulting in incomplete auditory scenes.
In contrast, our model with complete MoE Gating (top row) successfully integrates both speech and environmental sounds across on- and off-screen sources, producing coherent and holistic audio.

\section{Conclusion}
In this paper, we introduce Universal Holistic Audio Generation (UniHAGen), a new task that aims to synthesize complete auditory scenes encompassing both on-screen and off-screen sounds across diverse domains. To address UniHAGen, we propose OmniSonic, a flow-matching–based diffusion framework with a TriAttn-DiT backbone that jointly models multiple audio conditions. To support evaluation, we constructed UniHAGen-Bench, a benchmark covering three representative real-world scenarios. Experimental results demonstrate that OmniSonic achieves superior performance in both objective and subjective evaluations, effectively generating high-quality and semantically consistent audio across complex multi-source conditions.

\noindent
\textbf{Acknowledgments.} 
This work was supported in part by an Amazon Research Award. The opinions, findings, and conclusions expressed are those of the authors and do not necessarily reflect the views of Amazon.

{
    \small
    \bibliographystyle{ieeenat_fullname}
    \bibliography{main}
}

\clearpage

\appendix

\section{Appendix}

\subsection{Training and Inference Process}
\label{sec:method_training_inference}

\paragraph{Training.}
During the training of OmniSonic, for each data point, we randomly select one scenario from the three predefined configurations in our UniHAGen task: (1) Scenario 1: on-screen environmental sound + off-screen human speech, (2) Scenario 2: on-screen human speech + off-screen environmental sound, and (3) Scenario 3: on-screen environmental sound + off-screen environmental sound + off-screen human speech. 
For Scenario 1, we randomly select a video from our environmental audio–visual training set, \textit{i.e.}, VGGSound~\cite{chen2020vggsound}, to obtain the video clip with on-screen environmental sound and its corresponding caption. Then, we randomly sample a human speech audio from the speech training set (LRS3~\cite{afouras2018lrs3} and CommonVoice~\cite{ardila2020common}) to serve as the off-screen speech component, mixing it with the original video’s environmental audio according to a random signal-to-noise ratio (SNR) level. The resulting mixed waveform, along with its corresponding captions, transcriptions, and visual frames, is used as the training input for the model. 
For Scenario 2, we select a video from the speech audio–visual training set LRS3~\cite{afouras2018lrs3}, where the on-screen visual content corresponds to a speaking person. The associated speech and its transcription serve as the on-screen components. An environmental audio clip is then randomly sampled from the environmental training set and added as the off-screen component. The two sources are mixed at a randomly sampled SNR level, and the corresponding textual and visual conditions are used for conditioning the model.
For Scenario 3, we randomly select a video from the environmental audio–visual training set to provide the on-screen environmental sound and caption. An additional environmental audio clip and a human speech clip are then randomly sampled from their respective datasets to serve as the off-screen environmental sound and off-screen speech. All three sources are mixed using randomly sampled SNR levels, forming a complex multi-source auditory scene that mimics real-world conditions. The corresponding captions, transcriptions, and visual frames of each source type are used as conditioning inputs during training.

To enable classifier-free guidance (CFG)~\cite{ho2022classifier} during inference, we adopt a condition dropout strategy during training. Specifically, for each condition type, \textit{i.e.}, on-screen environmental caption, off-screen environmental caption, speech transcription, and visual frames, we randomly drop the entire condition with a specified probability, \textit{e.g.} 0.1. This strategy encourages the model to learn both conditional and unconditional generation behaviors, enhancing robustness to partially missing conditions and enabling controllable audio generation via CFG during inference.

To stabilize training and ensure high-quality speech generation, we adopt a two-stage training strategy. In the first stage, the model is trained only on speech data, where environmental sound–related conditions are kept empty (\textit{i.e.}, captions are empty strings). This stage helps the model effectively learn speech representation and synchronization without interference from environmental sound components. In the second stage, we switch to the full UniHAGen training setup described above, jointly learning to generate both speech and environmental sounds under different scenarios. This progressive training scheme prevents unstable optimization and improves the model’s ability to synthesize clear and coherent speech in complex multi-source audio scenes.

\paragraph{Inference.}
During inference, we adopt a multi-condition classifier-free guidance (CFG)~\cite{ho2022classifier,lee2024voiceldm} strategy to achieve controllable audio generation under different condition types. The modified velocity prediction is computed as:
\begin{equation}
    \begin{split}
        \tilde{\mathcal{V}}_\theta(\boldsymbol{x}_t, \boldsymbol{c}_{txt}^{on}, \boldsymbol{c}_{txt}^{off}, &\boldsymbol{c}_{txt}^{sp.}, \boldsymbol{c}_v) =  \mathcal{V}_\theta(\boldsymbol{x}_t, \boldsymbol{c}_{txt}^{on}, \boldsymbol{c}_{txt}^{off}, \boldsymbol{c}_{txt}^{sp.}, \boldsymbol{c}_v)) \\
        & + \lambda_{txt}^{on}( \mathcal{V}_\theta(\boldsymbol{x}_t, \boldsymbol{c}_{txt}^{on}, \varnothing, \varnothing, \boldsymbol{c}_v)) - \mathcal{V}_\theta' \\
        & + \lambda_{txt}^{off}( \mathcal{V}_\theta(\boldsymbol{x}_t, \varnothing, \boldsymbol{c}_{txt}^{off}, \varnothing, \boldsymbol{c}_v)) - \mathcal{V}_\theta') \\
        & + \lambda_{txt}^{sp.}( \mathcal{V}_\theta(\boldsymbol{x}_t, \varnothing, \varnothing, \boldsymbol{c}_{txt}^{sp.}, \boldsymbol{c}_v)) - \mathcal{V}_\theta')
    \end{split}
\end{equation}
where $\varnothing$ denotes the dropped (unconditioned) inputs, and $\mathcal{V}_\theta' = \mathcal{V}_\theta(\boldsymbol{x}_t, \varnothing, \varnothing, \varnothing, \boldsymbol{c}_v))$ represents the non-text unconditional prediction. The derivation is presented in Sec.~\ref{sec:derivation}.


\subsection{Derivation of Our Multi-Condition Classifier-Free Guidance}
\label{sec:derivation}
To extend classifier-free guidance (CFG)~\cite{ho2022classifier} to our four-condition setup, we consider the conditional distribution $p_{\boldsymbol{\theta}}(\boldsymbol{x}_t|\boldsymbol{c}_{1:3},\boldsymbol{c}_v)$, where $\boldsymbol{c}_1$, $\boldsymbol{c}_2$, $\boldsymbol{c}_3$ denote the three text-based conditions (on-screen environmental caption, off-screen environmental caption, and speech transcription), and $\boldsymbol{c}_v$ denotes the video condition. The flow model parameterizes this distribution through its time-dependent vector field. Following the modified CFG in~\cite{lee2024voiceldm} for dual conditions, we further enhance the influence of each conditioning signal by modifying the target conditional distribution. 

However, unlike the dual-condition case in~\cite{lee2024voiceldm}, the four conditions in our setting are not symmetric nor independent. The visual condition $\boldsymbol{c}_v$ is tightly coupled with one of the text-based conditions (the on-screen environmental caption or the speech transcription): visual frames directly reveal the on-screen sound source, its motion, and its temporal structure. As a result, the likelihood term $p_{\boldsymbol{\theta}}(\boldsymbol{c}_v|\boldsymbol{x}_t)$ is not independent from $p_{\boldsymbol{\theta}}(\boldsymbol{c}_1|\boldsymbol{x}_t)$ or $p_{\boldsymbol{\theta}}(\boldsymbol{c}_3|\boldsymbol{x}_t)$, where $\boldsymbol{c}_1$ and $\boldsymbol{c}_3$ denote conditions of on-screen environmental sound caption and speech transcription, respectively. Applying a CFG-style ``condition–unconditional" subtraction to $\boldsymbol{c}_v$ would therefore amplify shared information twice. This double-counting empirically leads to unstable guidance and degraded audio–visual consistency.

In contrast, the three text-based conditions $\boldsymbol{c}_1$, $\boldsymbol{c}_2$, $\boldsymbol{c}_3$ serve as independent semantic instructions: they specify what sound should occur (\textit{e.g.}, ``waves crashing", ``a dog barking", or speech content), but do not dictate how this sound temporally evolves with the visual scene. Thus, applying guidance to these three conditions is both well-defined and beneficial. Importantly, since the video condition provides the essential scene-level prior, we always retain $\boldsymbol{c}_v$ in both the conditional and the ``unconditional" branches, ensuring that the model never loses the scene context during guidance.

Under this formulation, the modified conditional distribution becomes:
\begin{equation}
    \begin{split}
        \tilde{p}_{\boldsymbol{\theta}}(\boldsymbol{x}_t|\boldsymbol{c}_{1:3},\boldsymbol{c}_v) \propto p_{\boldsymbol{\theta}}(\boldsymbol{x}_t|\boldsymbol{c}_{1:3},\boldsymbol{c}_v) \prod^3_{i=1}p_{\boldsymbol{\theta}}(\boldsymbol{c}_i|\boldsymbol{x}_t,\boldsymbol{c}_v)^{w_i}.
    \end{split}
\end{equation}
Taking the gradient of the log-density yields
\begin{equation}
    \begin{split}
        \nabla_{\boldsymbol{x}_t} \text{log}\,\tilde{p}_{\boldsymbol{\theta}}(\boldsymbol{x}_t|\boldsymbol{c}_{1:3},\boldsymbol{c}_v) &= \nabla_{\boldsymbol{x}_t} \text{log}\, p_{\boldsymbol{\theta}}(\boldsymbol{x}_t|\boldsymbol{c}_{1:3},\boldsymbol{c}_v) \\
        &+ \sum^3_{i=1} w_i \nabla_{\boldsymbol{x}_i}\text{log}\, p_{\boldsymbol{\theta}}(\boldsymbol{c}_i|\boldsymbol{x}_t,\boldsymbol{c}_v).
    \end{split}
    \label{eq:gradient}
\end{equation}
Using Bayes' rule, $p_{\boldsymbol{\theta}}(\boldsymbol{c}_i|\boldsymbol{x}_t,\boldsymbol{c}_v)=\frac{p_{\boldsymbol{\theta}}(\boldsymbol{x}_t|\boldsymbol{c}_i,\boldsymbol{c}_v)p(\boldsymbol{c}_i|\boldsymbol{c}_v)}{p_{\boldsymbol{\theta}}(\boldsymbol{x}_t|\boldsymbol{c}_v)}$, and noting that $p(\boldsymbol{c}_i|\boldsymbol{c}_v)$ does not depend on $\boldsymbol{x}_t$, we obtain
\begin{equation}
    \begin{split}
        \nabla_{\boldsymbol{x}_t}\text{log}\, p_{\boldsymbol{\theta}}(\boldsymbol{c}_i|\boldsymbol{x}_t,\boldsymbol{c}_v) = &\nabla_{\boldsymbol{x}_t}\text{log}\, p_{\boldsymbol{\theta}}(\boldsymbol{x}_t|\boldsymbol{c}_i,\boldsymbol{c}_v) \\ 
        &- \nabla_{\boldsymbol{x}_t}\text{log}\, p_{\boldsymbol{\theta}}(\boldsymbol{x}_t|\boldsymbol{c}_v).
    \end{split}
    \label{eq:bayes}
\end{equation}
Substituting Eq.~\ref{eq:bayes} into Eq.~\ref{eq:gradient}, we have:
\begin{equation}
    \begin{split}
        \nabla_{\boldsymbol{x}_t} \text{log}&\,\tilde{p}_{\boldsymbol{\theta}}(\boldsymbol{x}_t|\boldsymbol{c}_{1:3},\boldsymbol{c}_v) =
        \nabla_{\boldsymbol{x}_t}\text{log}\, p_{\boldsymbol{\theta}}(\boldsymbol{x}_t|\boldsymbol{c}_{1:3},\boldsymbol{c}_v) \\
        &+ \sum^3_{i=1} w_i (\nabla_{\boldsymbol{x}_t}\text{log}\,p_{\boldsymbol{\theta}}(\boldsymbol{x}_t|\boldsymbol{c}_i,\boldsymbol{c}_v) - \nabla_{\boldsymbol{x}_t}\text{log}\,p_{\boldsymbol{\theta}}(\boldsymbol{x}_t|\boldsymbol{c}_v)).
    \end{split}
    \label{eq:final}
\end{equation}
Finally, we rewrite Eq.~\ref{eq:final} in terms of the model’s predicted vector field, yielding the following guided velocity formulation:
\begin{equation}
    \begin{split}
        \tilde{\mathcal{V}}_{\boldsymbol{\theta}}(\boldsymbol{x}_t, \boldsymbol{c}_{1:3},\boldsymbol{c}_v) =& \mathcal{V}_{\boldsymbol{\theta}}(\boldsymbol{x}_t, \boldsymbol{c}_{1:3},\boldsymbol{c}_v) \\
        &+ \sum^3_{i=1}w_i(\mathcal{V}_{\boldsymbol{\theta}}(\boldsymbol{x}_t, \boldsymbol{c}_i,\boldsymbol{c}_v) - \mathcal{V}_{\boldsymbol{\theta}}(\boldsymbol{x}_t,\boldsymbol{c}_v)).
    \end{split}
\end{equation}

\subsection{Evaluation Metrics}

\begin{figure}[htbp]
    \centering
    \includegraphics[width=0.48\textwidth]{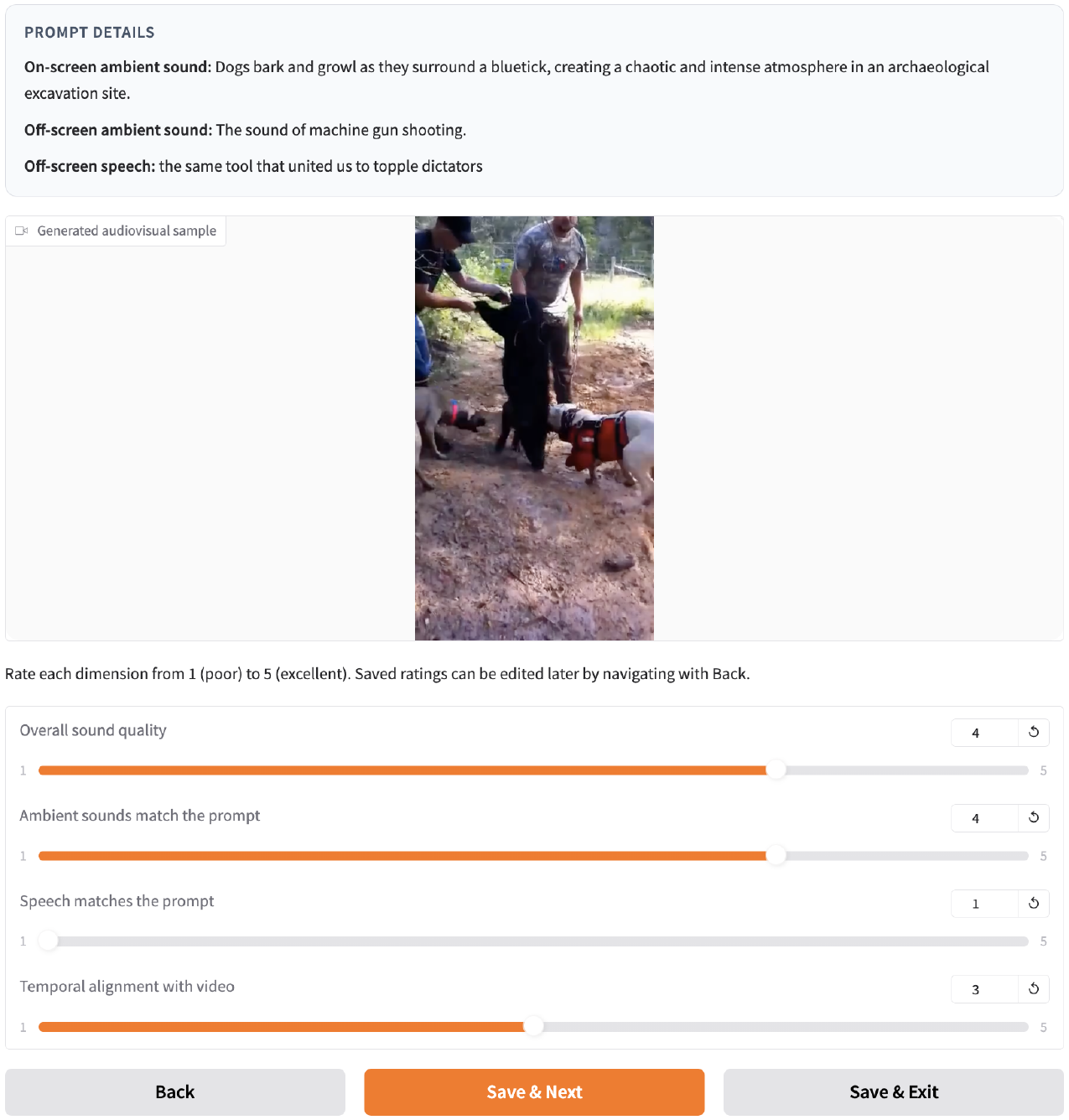}
    \caption{Interface for the subjective evaluation.}
    \label{fig:user_study_interface}
\end{figure}

We evaluate our OmniSonic and the baseline models using both objective and subjective evaluation metrics. For objective evaluation metrics, we adopt Fr{\'e}chet Audio Distance (FAD)~\cite{kilgour2018fr} and Mean Kullback–Leibler Divergence (MKL)~\cite{iashin2021taming} assess the perceptual quality and distributional similarity of generated audios. For semantic alignment evaluation, following previous works~\cite{kushwaha2025vintage, sheffer2023hear, wang2024v2a, zhang2024foleycrafter}, we use the AV score and AT score to measure the semantic correspondence between audio and video (AV) and between audio and text (AT), respectively. Specifically, we employ Wav2CLIP~\cite{wu2022wav2clip} to encode the generated audio into the CLIP~\cite{radford2021learning} feature space, enabling direct computation of cross-modal similarity with visual and textual embeddings. To evaluate the speech correctness in the generated audio, we utilize Word Error Rate (WER), Character Error Rate (CER), and Phoneme Error Rate (PER) to quantitatively assess the accuracy of synthesized speech content. Following~\cite{lee2024voiceldm}, we employ a pretrained Whisper~\cite{radford2023robust} model to transcribe the generated audio and compute the error rates by comparing the transcriptions with the ground-truth speech transcription. To measure audio–visual temporal synchronization, we adopt DeSync~\cite{cheng2025mmaudio,shan2025hunyuanvideo}, which utilizes Synchformer~\cite{iashin2024synchformer} to estimate the temporal misalignment between the generated audio and the corresponding video frames. For subjective evaluation, we conduct human listening tests to assess four aspects: overall quality (MOS-Q), environmental faithfulness (MOS-EF) for on-screen and/or off-screen environmental sounds, speech faithfulness (MOS-SF) for on-screen and/or off-screen speech, and temporal alignment (MOS-T) between the video and the on-screen sound. We randomly select 24 samples from our UniHAGen-Bench and generate the corresponding audio using OmniSonic and the compared baseline models. The all generated audios are randomly distributed among 13 human listeners, who rate them on a discrete 5-point scale. We report the mean opinion scores (MOS) averaged across all ratings for each evaluation aspect. The interface for this subjective evaluation is shown in Fig.~\ref{fig:user_study_interface}.

\subsection{Baselines}

We compare OmniSonic with state-of-the-art audio generation models: AudioLDM 2~\cite{liu2024audioldm}, VoiceLDM~\cite{lee2024voiceldm}, VinTAGe~\cite{kushwaha2025vintage}, MMAudio~\cite{cheng2025mmaudio}, and HunyuanVideo-Foley~\cite{shan2025hunyuanvideo}.
Among them, AudioLDM 2 is a general text-to-audio generation model. In addition to its official checkpoint, we also evaluate a speech-adapted version\footnote{\texttt{audioldm2-speech-gigaspeech checkpoint}} fine-tuned for speech-related tasks. VoiceLDM is a text-to-speech generation model designed to synthesize environmental speech, \textit{i.e.}, speech recorded or simulated within specific acoustic environments. VinTAGe is a recent advancement in joint video–text-to-audio generation for holistic auditory scene synthesis, aiming to produce both on-screen and off-screen sounds simultaneously. MMAudio and HunyuanVideo-Foley are video–text conditioned audio generation models built upon the MM-DiT~\cite{esser2024scaling} architecture, which leverages multimodal diffusion transformers to generate temporally aligned and semantically consistent audio from visual and textual cues.

\subsection{Implementation Details}

We implement OmniSonic in PyTorch~\cite{paszke2019pytorch}. For the model architecture, we use CLIP~\cite{radford2021learning} as the visual frame encoder, FLAN-T5~\cite{raffel2020exploring} as the environmental sound caption encoder, and SpeechT5~\cite{ao2022speecht5} as the speech transcription encoder. The differentiable Durator follows the design in~\cite{lee2024voiceldm,tan2024naturalspeech}, which consists of a Duration Predictor and a Learnable Upsampling Layer. The dimensions of the visual, speech transcription, and environmental sound caption embeddings are 512, 769, and 1024, respectively.
For the audio VAE, we adopt the pre-trained version from AudioLDM~\cite{liu2023audioldm}, which generate audio latent representation with number of channels of 8 and size of $256\times 16$. 
For the TriAttn-DiT, we set the patch size to 2 and the hidden size to 1152, and stack 28 blocks. 
We sample the audio at 16kHz, which is then transformed to spectrogram using Short-Time Fourier Transform (STFT) with FFT size of 1024, hop length of 160, and Hann window size of 1024. We pad or crop the waveform to a fixed duration of $1024\times 160$ before applying the STFT, yielding log-Mel spectrograms with $T=1024$ and $F=64$.
We train our model on 8 NVIDIA A100 GPUs with a global batch size of 64, the learning rate of 5e-5, and the weight decay of 1e-2. The model is optimized using the AdamW~\cite{loshchilov2018decoupled} optimizer. During training, the visual frame encoder and environmental sound caption encoder are frozen, while the speech transcription encoder and the differentiable Durator are trainable. 

For the CFG scales during sampling, we set different groups of values for the three different scenarios. Specifically, for scenario 1 (on-screen environmental sound + off-screen human speech), we set $\lambda^{on}_{txt}=5.0$, $\lambda^{sp.}_{txt}=2.5$, and $\lambda^{off}_{txt}=0.5$. For Scenario 2 (on-screen human speech + off-screen environmental sound), we set $\lambda^{sp.}_{txt}=7.5$, $\lambda^{off}_{txt}=2.5$, and $\lambda^{on}_{txt}=0.5$. For Scenario 3 (on-screen environmental sound + off-screen environmental sound + off-screen human speech), we set 
$\lambda^{on}_{txt}=5.0$, $\lambda^{off}_{txt}=2.5$, and $\lambda^{sp.}_{txt}=2.5$.


\subsection{Parameter Study}

We conduct parameter studies on the CFG scales. 
The results for Scenario 1 are presented in Tab.~\ref{tab:para_scenario_1}, where we fix $\lambda^{off}_{txt}=0.5$ and examine how varying $\lambda^{on}_{txt}$ and $\lambda^{sp.}_{txt}$ affects the performance.
We present the results of Scenario 2 in Tab.~\ref{tab:para_scenario_2}, where we fix $\lambda^{on}_{txt}=0.5$ and examine how varying $\lambda^{sp.}_{txt}$ and $\lambda^{off}_{txt}$ affects the performance.
The results of Scenario 3 is shown in Tab.~\ref{tab:para_scenario_3}, in which we investigate the impact of values of $\lambda^{on}_{txt}$, $\lambda^{off}_{txt}$, and $\lambda^{sp.}_{txt}$ on the final results.

\begin{table}[htbp]
  \centering
  \caption{Parameter study for Scenario 1 (on-screen environmental sound + off-screen human speech). We fix $\lambda^{off}_{txt}=0.5$ and examine the effect of varying $\lambda^{on}_{txt}$ and $\lambda^{sp.}_{txt}$ on the results.}
    \begin{tabular}{ccccc}
    \toprule
    $\lambda^{on}_{txt}$ & $\lambda^{sp.}_{txt}$ & FAD$\downarrow$   & Mean AV AT$\uparrow$ & WER$\downarrow$ \\
    \midrule
    5.0   & 2.5   & 3.40  & 19.62 & 0.15 \\
    5.0   & 7.5   & 5.39  & 19.50 & 0.14 \\
    7.5   & 2.5   & 4.14  & 19.02 & 0.16 \\
    7.5   & 5.0   & 4.98  & 19.32 & 0.15 \\
    7.5   & 7.5   & 5.04  & 19.28 & 0.15 \\
    9.5   & 2.5   & 4.23  & 18.94 & 0.17 \\
    9.5   & 5.0   & 5.19  & 19.32 & 0.15 \\
    9.5   & 7.5   & 5.40  & 19.30 & 0.15 \\
    12.5  & 2.5   & 4.06  & 18.64 & 0.18 \\
    12.5  & 5.0   & 4.94  & 18.82 & 0.16 \\
    12.5  & 7.5   & 5.39  & 18.95 & 0.16 \\
    \bottomrule
    \end{tabular}%
  \label{tab:para_scenario_1}%
\end{table}%

\begin{table}[htbp]
  \centering
  \caption{Parameter study for Scenario 2 (on-screen human speech + off-screen environmental sound). We fix $\lambda^{on}_{txt}=0.5$ and examine the effect of varying $\lambda^{sp.}_{txt}$ and $\lambda^{off}_{txt}$ on the results.}
    \begin{tabular}{ccccc}
    \toprule
    $\lambda^{sp.}_{txt}$ & $\lambda^{off}_{txt}$ & FAD   & Mean AV AT & WER \\
    \midrule
    2.5   & 2.5   & 2.63  & 18.13 & 0.11 \\
    2.5   & 5.5   & 4.77  & 17.26 & 0.12 \\
    2.5   & 7.5   & 5.32  & 17.19 & 0.10 \\
    5.5   & 2.5   & 2.33  & 17.95 & 0.12 \\
    5.5   & 5.5   & 4.42  & 17.55 & 0.11 \\
    5.5   & 7.5   & 5.36  & 17.18 & 0.11 \\
    7.5   & 2.5   & 2.59  & 17.95 & 0.12 \\
    7.5   & 5.5   & 4.79  & 17.51 & 0.11 \\
    7.5   & 7.5   & 5.09  & 17.45 & 0.12 \\
    \bottomrule
    \end{tabular}%
  \label{tab:para_scenario_2}%
\end{table}%

\begin{table}[htbp]
  \centering
  \caption{Parameter study for Scenario 3 (on-screen environmental sound + off-screen environmental sound + off-screen human speech). We examine the effect of varying $\lambda^{on}_{txt}$, $\lambda^{off}_{txt}$, and $\lambda^{sp.}_{txt}$ on the results.}
    \begin{tabular}{cccccc}
    \toprule
    $\lambda^{on}_{txt}$ & $\lambda^{off}_{txt}$ & $\lambda^{sp.}_{txt}$ & FAD$\downarrow$   & Mean AV AT$\uparrow$ & WER$\downarrow$ \\
    \midrule
    5.0   & 2.5   & 2.5   & 3.39  & 18.26 & 0.16 \\
    5.0   & 2.5   & 3.5   & 3.97  & 18.55 & 0.15 \\
    5.0   & 5.0   & 2.5   & 3.66  & 17.00 & 0.16 \\
    5.0   & 5.0   & 3.5   & 3.98  & 17.50 & 0.14 \\
    7.5   & 2.5   & 2.5   & 3.88  & 18.30 & 0.15 \\
    7.5   & 2.5   & 3.5   & 4.34  & 18.54 & 0.18 \\
    7.5   & 5.0   & 2.5   & 3.70  & 17.48 & 0.15 \\
    7.5   & 5.0   & 3.5   & 4.07  & 17.64 & 0.15 \\
    7.5   & 5.0   & 5.0   & 4.95  & 17.70 & 0.13 \\
    \bottomrule
    \end{tabular}%
  \label{tab:para_scenario_3}%
\end{table}%

\subsection{Limitations and Future Work}

Although OmniSonic achieves strong performance across diverse mixed-source scenarios, several limitations remain.

First, the training samples used in our UniHAGen task are synthetically constructed by combining audio, text, and video clips from VGGSound, LRS3, and CommonVoice to simulate the three scenario configurations. While this composition strategy enables controlled supervision across on/off-screen speech–environment combinations, it does not fully capture the richness, spontaneity, and acoustic complexity of truly in-the-wild audio–visual scenes. Moreover, synthetic mixing often results in acoustic inconsistencies, such as mismatched loudness, differing recording conditions, or unnatural blending between speech and environmental sounds, which limits the realism of the training distribution.

Future work may explore collecting large-scale natural audio–visual corpora with organically co-occurring speech and environmental events, or developing more advanced simulation pipelines to better approximate real-world multimodal dynamics.

Second, our model relies solely on CLIP visual features for video conditioning. Although CLIP provides strong global semantic understanding, it lacks fine-grained temporal sensitivity. In visually stable or weakly dynamic scenes, where consecutive frames exhibit minimal variation, CLIP features tend to remain nearly invariant, limiting the model’s ability to infer subtle temporal cues for precise audio–visual synchronization. This leads to weaker performance on synchronization-focused metrics compared with models that incorporate temporally specialized encoders such as Synchformer. 

Future work may integrate more temporally expressive video representations or design dedicated audio-aware video encoders to strengthen fine-grained synchronization without compromising semantic grounding.


\end{document}